\documentclass[sigconf]{acmart}

\usepackage{algorithmic}
\usepackage{graphicx}
\usepackage{textcomp}
\usepackage{xcolor}
\usepackage{multirow}
\usepackage{bm}
\usepackage{framed}

\usepackage{listings, xcolor}

\definecolor{verylightgray}{rgb}{.97,.97,.97}

\lstdefinelanguage{Solidity}{
	keywords=[1]{anonymous, assembly, assert, balance, break, call, callcode, case, catch, class, constant, continue, constructor, contract, debugger, default, delegatecall, delete, do, else, emit, event, experimental, export, external, false, finally, for, function, gas, if, implements, import, in, indexed, instanceof, interface, internal, is, length, library, log0, log1, log2, log3, log4, memory, modifier, new, payable, pragma, private, protected, public, pure, push, require, return, returns, revert, selfdestruct, send, solidity, storage, struct, suicide, super, switch, then, this, throw, transfer, true, try, typeof, using, value, view, while, with, addmod, ecrecover, keccak256, mulmod, ripemd160, sha256, sha3}, 
	keywordstyle=[1]\color{blue}\bfseries,
	keywords=[2]{address, bool, byte, bytes, bytes1, bytes2, bytes3, bytes4, bytes5, bytes6, bytes7, bytes8, bytes9, bytes10, bytes11, bytes12, bytes13, bytes14, bytes15, bytes16, bytes17, bytes18, bytes19, bytes20, bytes21, bytes22, bytes23, bytes24, bytes25, bytes26, bytes27, bytes28, bytes29, bytes30, bytes31, bytes32, enum, int, int8, int16, int24, int32, int40, int48, int56, int64, int72, int80, int88, int96, int104, int112, int120, int128, int136, int144, int152, int160, int168, int176, int184, int192, int200, int208, int216, int224, int232, int240, int248, int256, mapping, string, uint, uint8, uint16, uint24, uint32, uint40, uint48, uint56, uint64, uint72, uint80, uint88, uint96, uint104, uint112, uint120, uint128, uint136, uint144, uint152, uint160, uint168, uint176, uint184, uint192, uint200, uint208, uint216, uint224, uint232, uint240, uint248, uint256, var, void, ether, finney, szabo, wei, days, hours, minutes, seconds, weeks, years},	
	keywordstyle=[2]\color{teal}\bfseries,
	keywords=[3]{block, blockhash, coinbase, difficulty, gaslimit, number, timestamp, msg, data, gas, sender, sig, value, now, tx, gasprice, origin},	
	keywordstyle=[3]\color{violet}\bfseries,
	identifierstyle=\color{black},
	sensitive=false,
	comment=[l]{//},
	morecomment=[s]{/*}{*/},
	commentstyle=\color{gray}\ttfamily,
	stringstyle=\color{red}\ttfamily,
	morestring=[b]',
	morestring=[b]"
}

\AtBeginDocument{%
  \providecommand\BibTeX{{%
    \normalfont B\kern-0.5em{\scshape i\kern-0.25em b}\kern-0.8em\TeX}}}

\begin{document}

\title{Semantic Code Search for Smart Contracts}
\author{Chaochen Shi}
\affiliation{%
 \institution{Deakin University}
 \country{Australia}}
\email{shicha@deakin.edu.au}

\author{Yong Xiang}
\affiliation{%
 \institution{Deakin University}
 \country{Australia}}
\email{yong.xiang@deakin.edu.au}

\author{Jiangshan Yu}
\affiliation{%
 \institution{Monash University}
 \country{Australia}}
\email{jiangshan.yu@monash.edu}

\author{Longxiang Gao}
\affiliation{%
 \institution{Deakin University}
 \country{Australia}}
\email{longxiang.gao@deakin.edu.au}
\begin{abstract}
Semantic code search technology allows searching for existing code snippets through natural language, which can greatly improve programming efficiency. Smart contracts, programs that run on the blockchain, have a code reuse rate of more than 90\%, which means developers have a great demand for semantic code search tools. However, the existing code search models still have a semantic gap between code and query, and perform poorly on specialized queries of smart contracts. In this paper, we propose a Multi-Modal Smart contract Code Search (MM-SCS) model. Specifically, we construct a Contract Elements Dependency Graph (CEDG) for MM-SCS as an additional modality to capture the data-flow and control-flow information of the code. To make the model more focused on the key contextual information, we use a multi-head attention network to generate embeddings for code features. In addition, we use a fine-tuned pretrained model to ensure the model's effectiveness when the training data is small. We compared MM-SCS with four state-of-the-art models on a dataset with 470K (code, docstring)  pairs collected from Github and Etherscan. Experimental results show that MM-SCS achieves an MRR (Mean Reciprocal Rank) of 0.572, outperforming four state of-the-art models UNIF, DeepCS, CARLCS-CNN, and TAB-CS by 34.2\%, 59.3\%, 36.8\%, and 14.1\%, respectively. Additionally, the search speed of MM-SCS is second only to UNIF, reaching 0.34s/query.
\end{abstract}

\ccsdesc[300]{Software and its engineering~Search-based software engineering}
\keywords{code search, smart contract, attention mechanism, graph representation}

\maketitle

\section{Introduction}
The smart contract on blockchain is an auto-running digital protocol. By coding smart contracts, developers can construct personalized decentralized applications (DAPPs) with a high-level programming language Solidity. Over the recent years, DAPPs on blockchain platforms represented by Ethereum have proliferated. As of 28th August 2021, the number of smart contracts deployed on Ethereum had reached 1.51 million, so that a double growth rate had been realized as compared to the beginning of 2021. The communal ecology where Ethereum is active has attracted numerous developers to put their ideas into practice by developing smart contracts. The code search engine is a vital tool to boost the programming efficiency of smart contract developers. It allows developers to search existing code snippets from a large-sized code repository for further reference or reuse. The code search methods provided by the current mainstream code repositories (e.g. Github, Etherscan) are based on tag or keyword search. However, quite a lot of work is underway to explore semantic code search, a more natural code search method.

Semantic code search is a technique that operates by searching in natural language from the code repository and returning the code snippets that consist with query semantics. Early code search techniques typically deemed the code as a text and compared the text similarity between code and query using the information retrieval model~\cite{DBLP:journals/sigir/Can93}. This method is lacking in the ability of capturing the semantics of code and query, hence inferior in performance. The latest research has started to adopt neural networks -- a method referred to as neural code search -- to build code search engines. To bridge the gap between programming languages and natural languages, neural code search typically maps code and query into a shared vector space and measures their semantic similarity via vector distance. Therefore, the key challenges to neural code search are how to capture the semantics of code and query and generate the accurate embedding.

There have been a few representative cutting-edge technologies. NCS~\cite{DBLP:conf/pldi/SachdevLLKS018} is an unsupervised learning technology. It generates word embeddings for code and query tokens through a shared fastText~\cite{DBLP:journals/tacl/BojanowskiGJM17} encoder and generates sentence- or document-level embeddings according to TF-IDF weights. However, this method depends on the overlapped words in the code snippet and query. If the query contains a word that does not exist in the code corpus, the accuracy of NCS model will be reduced significantly. The supervised learning based model DeepCS~\cite{DBLP:conf/icse/GuZ018} has solved this problem. DeepCS operates by learning the embeddings of code and query via two Long Short-Term Memory (LSTM)~\cite{hochreiter1997long} networks, respectively, and comparing them in the shared vector space. Most of the current state-of-the-art models have been improved based on DeepCS, introduced in Section 2.1. However, the performance and applicability of these DeepCS based models are limited due to two key facts. First, the models are hard to capture the deep dependency between the key code elements. The textual features of these models include code tokens, function names, API sequence, even the traversal of Abstract Syntax Tree (AST). However, some types of structural information of the code, such as control- and data-flow, are lacking support to capture, thereby losing partial semantics of code. This greatly limits its performance in terms of search accuracy. Second, such models require large amounts of (code, query) pairs as a corpus for training, and consume considerable computing resources to achieve an ideal performance. This limits its applicability to other applications. For example, there remains no large public corpus of Solidity analogous to CodeSearchNet~\cite{DBLP:journals/corr/abs-1909-09436} for training at present, making it challenging to adopt such algorithms in blockchain applications.

This paper proposes a model called Multi-Modal Smart contract Code Search (MM-SCS) to address the limitations mentioned above. Specifically, MM-SCS improves neural code search in three aspects:

\textbf{Extra modality}. There have been works~\cite{wang2020detecting} that use control- and data-flow to represent code structure in program vulnerability detection and code comprehension~\cite{DBLP:conf/nips/Ben-NunJH18} tasks. Based on the hypothesis that control- and data-flow can also help leverage the code structural information in semantic search, we propose the concept of Contract Elements Dependency Graph (CEDG) as an extra modality explained in Section 3. CEDG integrates both control- and data-flow information in a single graph. Compared with AST and code property graph~\cite{DBLP:conf/sp/YamaguchiGAR14}, CEDG further highlights the dependency between elements of code while becoming simpler-structured, which is beneficial to the learning of key semantic features. Moreover, CEDG considers the differences between Solidity and other programming languages. For example, Solidity have a unique programmable fallback function with no name and no argument. If the contract is called without matching any function or passing any data, the fallback function would be triggered. Besides, Solidity has multiple unique keywords and statements. For example, the keyword \textsf{modifier} acts on a function, and the logic in the modifier can be pre-executed either before or after the function; the exception handling statement \textsf{require()} defines the conditions that the function needs to meet to continue execution, etc. There are specific nodes and edges designed for such characteristics of Solidity in CEDG. Although CEDG is specificlly designed for Solidity, the general idea of considering code elements dependency can also be applied to other programming languages. 

\textbf{Code embedding mechanisms}. MM-SCS adopts multi-head self-attention networks~\cite{DBLP:conf/nips/VaswaniSPUJGKP17} to embed three textual modalities (code tokens, function name, and API sequence), and a modified graph attention network~\cite{DBLP:conf/acl/NathaniCSK19} to embed CEDG, respectively, introduced in Section 4.2. Compared with existing models with no attention mechanism~\cite{DBLP:conf/icse/GuZ018} or with only single headed attention networks~\cite{DBLP:conf/sigsoft/CambroneroLKS019, DBLP:conf/iwpc/ShuaiX0Y0L20, DBLP:conf/wcre/XuYLSYL021}, multi-head attention mechanism can effectively learn different features from different heads and emphasize important features.

\textbf{Pretained model}. Husain~\cite{husain2019codesearchnet} highlighted an open challenge about measuring whether pretrained models are useful in code search tasks. In MM-SCS, we adopt the fine-tuned pretrained model ALBERT as the encoder of queries as in Section 4.3 to address this challenge. Our target is to achieve the optimal performance under the limited training data with the help of ALBERT's large corpora.

To evaluate MM-SCS, we collected approximately 470K annotated smart contract snippets written in Solidity from Github and Etherscan.io and sorted them into (code, docstring) pairs. After deduplication, there are 48K records remaining. The experimental results in Section 5 demonstrate MM-SCS outperforms four state of-the-art models UNIF~\cite{DBLP:conf/sigsoft/CambroneroLKS019}, DeepCS~\cite{DBLP:conf/icse/GuZ018}, CARLCS-CNN~\cite{DBLP:conf/iwpc/ShuaiX0Y0L20}, and TAB-CS~\cite{DBLP:conf/wcre/XuYLSYL021} by 34.2\%, 59.3\%, 36.8\%, and 14.1\% in terms of MRR, respectively. In addition, MM-SCS achieves 0.34s/query in the testing process, which is only second to UNIF. 

The main contributions of this paper are listed as follows:
\begin{itemize}
\item We propose the model MM-SCS dedicated to smart contract semantic code search, which improves neural code search with extra modality, code embedding mechanisms and pretrained model. Experimental results in Section 5 show MM-SCS outperforms four state-of-the-art models as baselines by at least 14.1\% in terms of MRR.
\item We put forward CEDG, a novel graph representation of code which integrates dependency between code elements including control- and data-flow, as an extra modality used in MM-SCS. Althrough CEDG is designed for Solidity in this paper, the general idea is independent and can be applied to other programming languages.
\item We build a corpus with originally 470K (code, docstring) pairs for experiments. The corpus will be made public to promote research in the field of smart contract code search.
\end{itemize}

\section{Background and related work}
\subsection{Neural Code Search}
As mentioned in Section 1, neural code search falls mainly into unsupervised learning and supervised learning methods. A representative unsupervised learning method is NCS proposed by Facebook team~\cite{DBLP:conf/pldi/SachdevLLKS018}. NCS merely uses the tokens from the code corpus for word embedding and highly depends on the overlapped words between code and query. UNIF~\cite{DBLP:conf/sigsoft/CambroneroLKS019} is a supervised extension to NCS. It uses a bag-of-words-based network to learn the mapping from code tokens into query tokens and introduces the TF-IDF weight based attention mechanism. DeepCS~\cite{DBLP:conf/icse/GuZ018} generates joint embedding from code tokens, function names and API sequence, and learns the code-query relation via LSTM. Recent works concentrated on the improvement of DeepCS. For example, the method CARLCS-CNN~\cite{DBLP:conf/iwpc/ShuaiX0Y0L20} introduces a co-attentive mechanism on the base of DeepCS architecture; Yan. et al.~\cite{DBLP:conf/wcre/XuYLSYL021} designed a two-stage-attention-based model for their TAB-CS model. Such attention-based models have improved the primordial DeepCS to better capture the long-range relationship between tokens, thereby achieving a performance superior to DeepCS.

\begin{figure}[t]
\centerline{\includegraphics[width=0.5\textwidth]{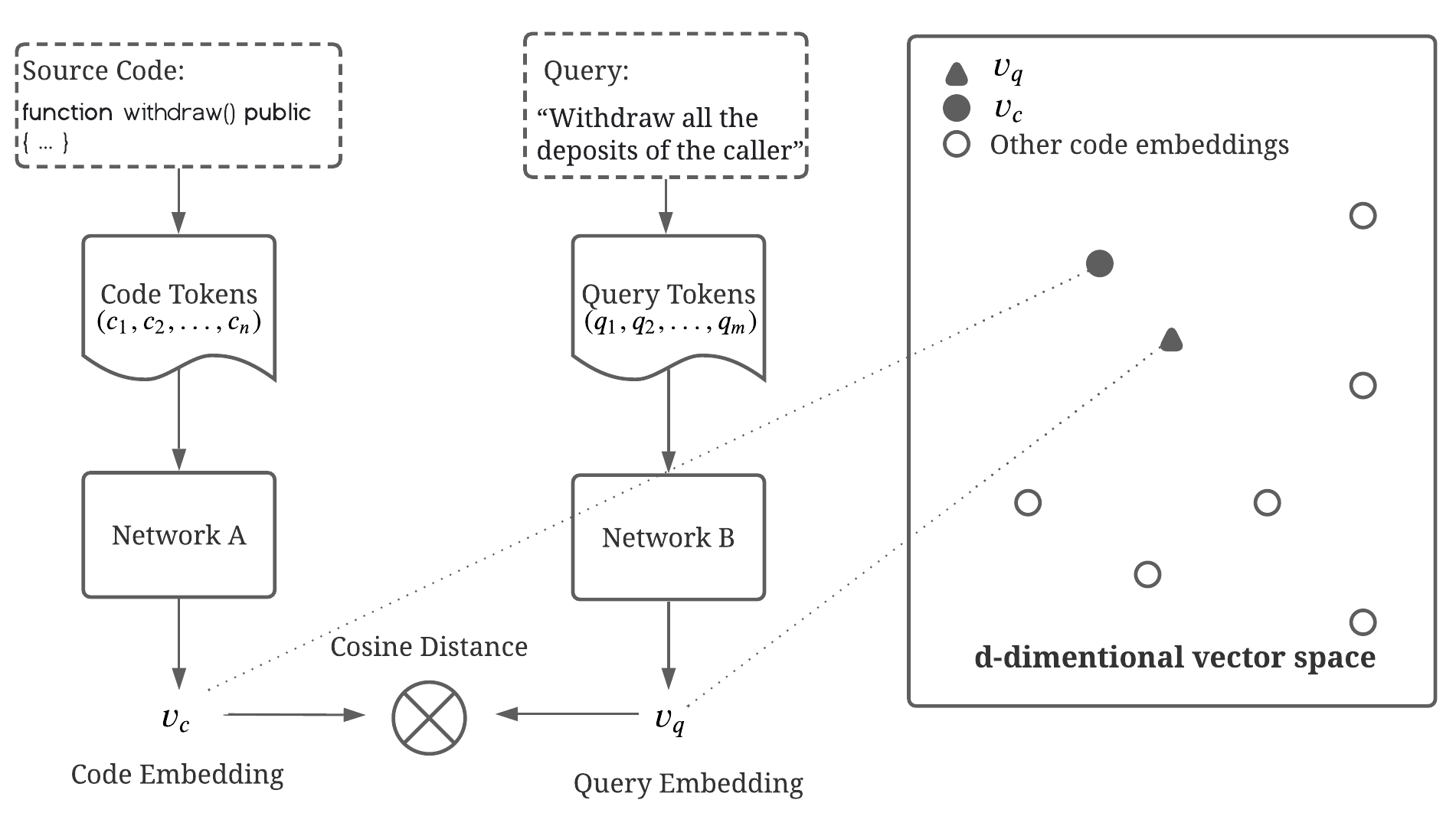}}
\caption{General architecture of neural code search}
\label{fig1}
\end{figure}

All the above-mentioned neural code search methods comply with the architecture of pseudo-siamese network~\cite{DBLP:conf/sigsoft/CambroneroLKS019}. As shown in Fig.~\ref{fig1}, two subnetworks of pseudo-siamese network each receive a tokenized sequence $c_1,\ldots,c_n$ and $q_1,\ldots,q_n$ from code and query and output the corresponding embedding $v_c \in \mathbb{R}^d$ and $v_q \in \mathbb{R}^d$, respectively. The similarity between the two inputs can be derived from the distance between $v_c$ and $v_q$ calculated by Eq.~(\ref{eq1}). The greater the value of the cosine distance between them, the closer the semantics represented by them. Therefore the search result can be returned by sorting the values of cosine distance.
\begin{equation}
\label{eq1}
cos(v_c, v_q) =\frac{v_c \cdot v_q}{ \|{v_c}\| \cdot \|{v_q}\|}
\end{equation}

\subsection{Sequence Embedding}
In the context of neural network, embedding is a mapping process from discrete variables into continuous vectors. One typical word embedding technique is word2vec~\cite{DBLP:journals/corr/abs-1301-3781}, which utilizes Continuous Bag of Words Model (CBOW) or Skip-gram model to map words into vectors and measures their semantic similarity using the vector distance. By concatenating the vectors of single words in a sentence, the sequence of words can be mapped into a vector that represents semantics of the entire sentence~\cite{DBLP:journals/corr/PalangiDSGHCSW15}.

In neural code search, we need to embed the input sequences from code and query sentences, respectively. For code information, UNIF only receives the code token sequences as the input. DeepCS receives three inputs: code token, function name, and API sequence. Among them, code tokens are embedded as $v_{token}$ via a Multi Layer Perceptron (MLP), whereas function names and API sequences are embedded as $v_{name}$ and $v_{API}$, respectively, via LSTM. The final output code embedding $v_c$ is as shown in Eq.~(\ref{eq3}). CARLCS-CNN has replaced MLP and $LSTM_1$ with CNN, as in Eq.~(\ref{eq4}), while TAB-CS has replaced all code encoders with attention networks, as in Eq.~(\ref{eq5}).
\begin{equation}
\label{eq3}
v_c=MLP(v_{token})+LSTM_1(v_{name})+LSTM_2(v_{API})
\end{equation}

\begin{equation}
\label{eq4}
v_c=CNN_1(v_{token})+CNN_2(v_{name})+LSTM(v_{API})
\end{equation}

\begin{equation}
\label{eq5}
v_c=Att_1(v_{token})+Att_2(v_{name})+Att_3(v_{API})
\end{equation}

Query is a natural language sentence made up of word sequences, hence it can also be mapped into the vector space with the exact dimensions as code embedding. DeepCS and CARLCS-CNN generate embedding $v_q$ of query sentence using LSTM and CNN, respectively, while UNIF and TAB-CS use attention network as the encoder.

\subsection{Graph Embedding}
Graph embedding is a method to map the features of nodes and edges of a graph into a low-dimensional vector space. Due to the complexity of graph structure, a major challenge of graph embedding is how to save as integral as possible the network topology and node information, so as to achieve a better performance in downstream tasks like deep learning. DeepWalk~\cite{DBLP:conf/kdd/PerozziAS14} is a typical graph embedding method which learns the partial representation of nodes in a graph through truncated random walk. Node2vec~\cite{DBLP:conf/kdd/GroverL16} is an extension to DeepWalk, which uses skip-gram to adjust the weight of random walk. Graph convolutional network (GCN)~\cite{DBLP:conf/iclr/KipfW17} expands convolution operation from traditional data (e.g. image) to graph data. Its core idea is to perform first-order approximation on the convolution kernel around each node. Graph attention network (GAT)~\cite{velivckovic2017graph} introduces the self-attention mechanism on the base of GCN. The advantage is that any size of input can be processed, with attention to the most relevant portions in topology.

In training, the above-mentioned models lay an emphasis on embedding node features while overlooking edge features. In a knowledge graph, edges typically contain rich features, with significant contributions to the semantics of the entire graph. Nathani~\cite{DBLP:conf/acl/NathaniCSK19} proposed a GAT based method for knowledge graph. This method captures both node features and edge(relation) features in a knowledge graph and exhibits a superior performance to the predecessors in relation prediction tasks. Similar to  knowledge graph, edges play an important role of expressing program semantics in the CEDG proposed in this paper. Therefore, we adopt the method proposed by Nathani to generate embedding of CEDG. The experiment evinces this method has achieved a satisfactory performance in smart contract code search tasks.

\section{contract elements dependency graph}
\subsection{Node Representation}
Contract elements are classified into three categories of nodes: Invocation, Variable, and Fallback nodes. Invocation nodes represent the function invocations in the contract, including self-defined functions and the system functions of Solidity such as \textsf{transfer()}, \textsf{send()}, \textsf{call()}, etc. Specifically, the function definition is also counted as a special invocation, so as to refer to itself when the function name appears for the first time. Variable nodes represent the variables in the contract, including self-defined variables and system variables, such as \textsf{msg.sender} which refers to the address of the external caller. Fallback nodes refer to the fallback function, which is a special function of smart contracts~\cite{Sol}. It is executed when a function not existing within the contract is being invoked or when a function is being invoked without providing data. The fallback function is devoid of name or parameter(s), and only one such can be defined within a contract.

\begin{figure}[t]
\centerline{\includegraphics[width=0.5\textwidth]{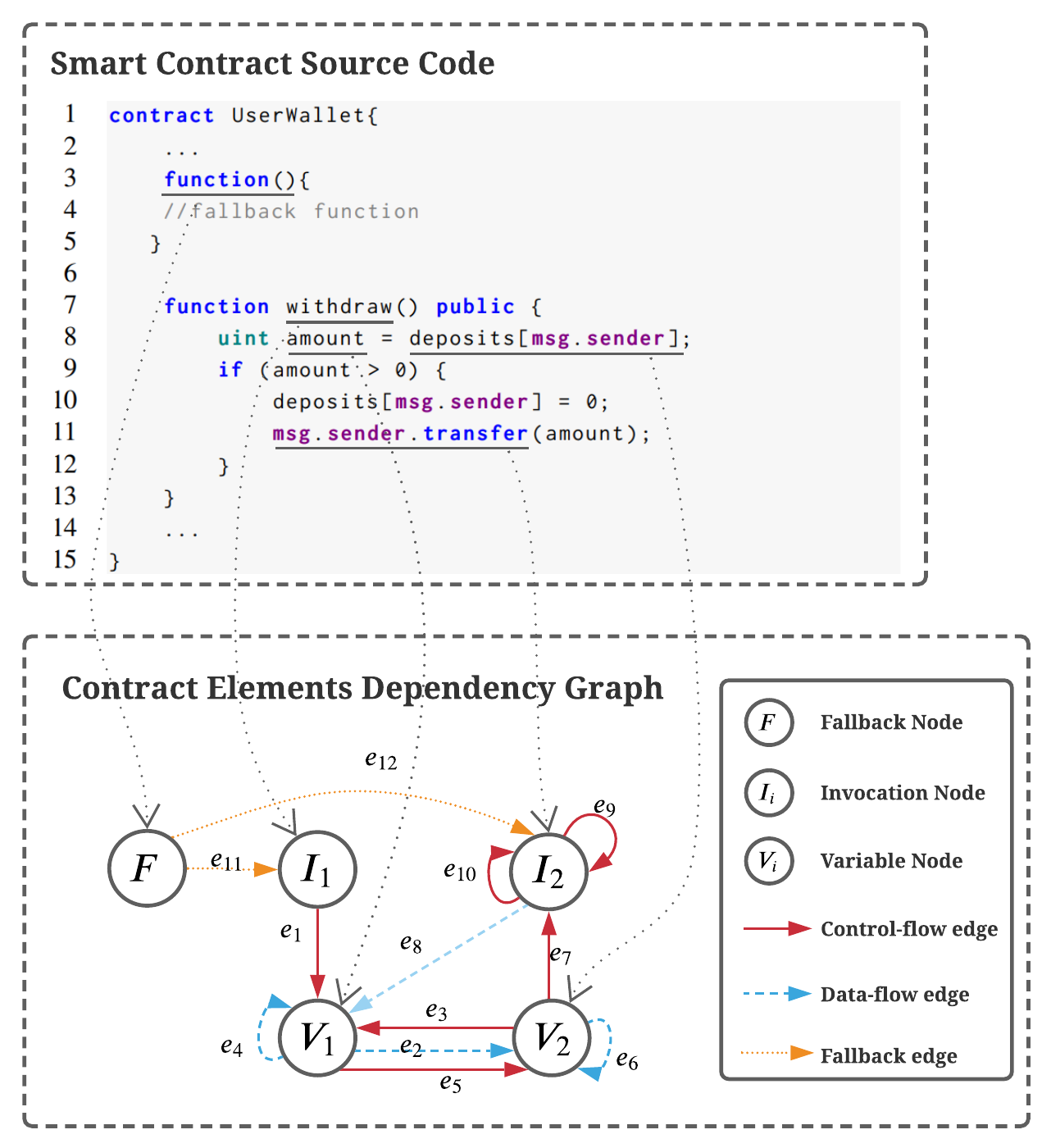}}
\caption{An example of Contract Elements Dependency Graph.}
\label{fig2}
\end{figure}

The attributes of nodes are [Category, Type, Name], which are obtained by traversing through the AST. Among the attributes, Category is one of the three categories of nodes, and Type is the type of a function, modifier or variable. Depending on the definition of Solidity documents, the type of function is classified into internal, external, public or private, and the type of modifier is consistent with that of function; the type of variable may be uint, address, and so forth. Name among the attributes may be either function name or variable name. In Fig.~\ref{fig2}, for example, $I_1$ refers to the function \textsf{withdraw()}, whose attributes are [Invocation, internal, withdraw]; $I_2$ refers to the variable \textsf{amount}, whose attributes are [Variable, uint, amount]. Fallback node is devoid of name, and so we set its type as fallback and name as 0.

\subsection{Edge Representation}
Enlightened by the contract graph for vulnerability detection in~\cite{zhuang2020smart}, we have defined different types of edges to represent the control-flow, data-flow and fallback relationship between the elements of contract. See Table 1 for the specific types of edges, where the control-flow edge corresponds to the conditional, loop and exception handling statements in code. Specifically, we add the edges corresponding to the start and end of blocks in order to represent the code blocks in the source code. The data-flow edge corresponds to the usage of variables. The fallback edge represents the relation between fallback nodes and the nodes that may trigger the fallback function. The attributes of edges are [$V_s$,$V_e$,Type,Order], which are obtained through pattern matching among code. Where $V_s$ and $V_e$ represent the start point and end point of an edge, respectively. For the relation between neighboring statements, $V_s$ refers to the last element in the anterior statement, whereas $V_e$ refers to the first element in the posterior statement. For example, $V_s$ and $V_e$ of $e_3$ in Fig.~\ref{fig2} correspond respectively to \textsf{deposits[msg.sender]} in code line 8 and \textsf{amount} in line 9. Attribute Type is the type of the edges shown in Table 1. Attribute Order represents the order in which the edges appear, namely the time sequence information during execution of the code. See Fig.~\ref{fig2} and Table 2 for examples of edges.

\begin{table}[]
\centering
\caption{Edges Defined in CEDG}
\label{tab1}
\begin{tabular}{|c|c|c|}
\hline
\textbf{Category}             & \textbf{Type(Abbr.)} & \textbf{Semantic Fact}          \\ \hline
\multirow{11}{*}{Control-flow} & IF                          & if (...) then \{...\}           \\
                              & IE                          & if (...) else \{...\}           \\
                              & WH                          & while (...) do\{...\}           \\
                              & FR                          & for (...) do\{...\}             \\
                              & TC                          & try \{...\} catch \{...\}       \\
                              & AT                          & assert(...) 
                              \\
                              & RT                          & revert(...)                       \\
                              & RQ                          & require(...)                      \\
                              & BS                          & block start                     \\
                              & BE                          & block end                       \\
                              & NS                          & natural sequential relationship \\ \hline
\multirow{2}{*}{Data-flow}    & AS                          & assgin operation                \\
                              & AC                          & access operation                \\ \hline
Fallback                      & FB                          & fallback relationship           \\ \hline
\end{tabular}
\end{table}

\begin{table}[]
\centering

\label{tab2}
\caption{Edges Shown in Fig. 2}
\scalebox{0.9}{
\begin{tabular}{|c|c|c|c|c|c|c|c|c|c|c|c|c|}
\hline
\multicolumn{1}{|l|}{}     & \bm{$e_1$}  & \bm{$e_2$} & \bm{$e_3$} & \bm{$e_4$} &  \bm{$e_5$}& \bm{$e_6$} & \bm{$e_7$} & \bm{$e_8$} & \bm{$e_9$} & \bm{$e_{10}$} & \bm{$e_{11}$} & \bm{$e_{12}$}   \\ \hline
\bm{$V_s$} & $I_1$                       & $V_1$                       & $V_2$                              & $V_1$                                & $V_1$ & $V_2$ & $V_2$ & $I_2$ & $I_2$ & $I_2$ & $F$ & $F$ \\ \hline
\bm{$V_e$}   & $V_1$                       & $V_2$                       & $V_1$                              & $V_1$                                & $V_2$ & $V_2$ & $I_2$ & $V_1$  & $I_2$ &$I_2$  & $I_1$ & $I_2$ \\ \hline
\textbf{Type}  & BS                       & AS                      & IF  
                           & AC                                & BS & AS &NS  & AC & BE & BE & FB &  FB\\ \hline

\textbf{Order}   & 1                       & 2                       & 3                             & 4                               & 5 & 6 & 7  & 8 & 9 & 10 & 11 & 12 \\ \hline

\end{tabular}}
\end{table}

\section{Proposed semantic code search approach}
\subsection{Overview}
\begin{figure*}[htbp]
\centerline{\includegraphics[width=0.7\textwidth]{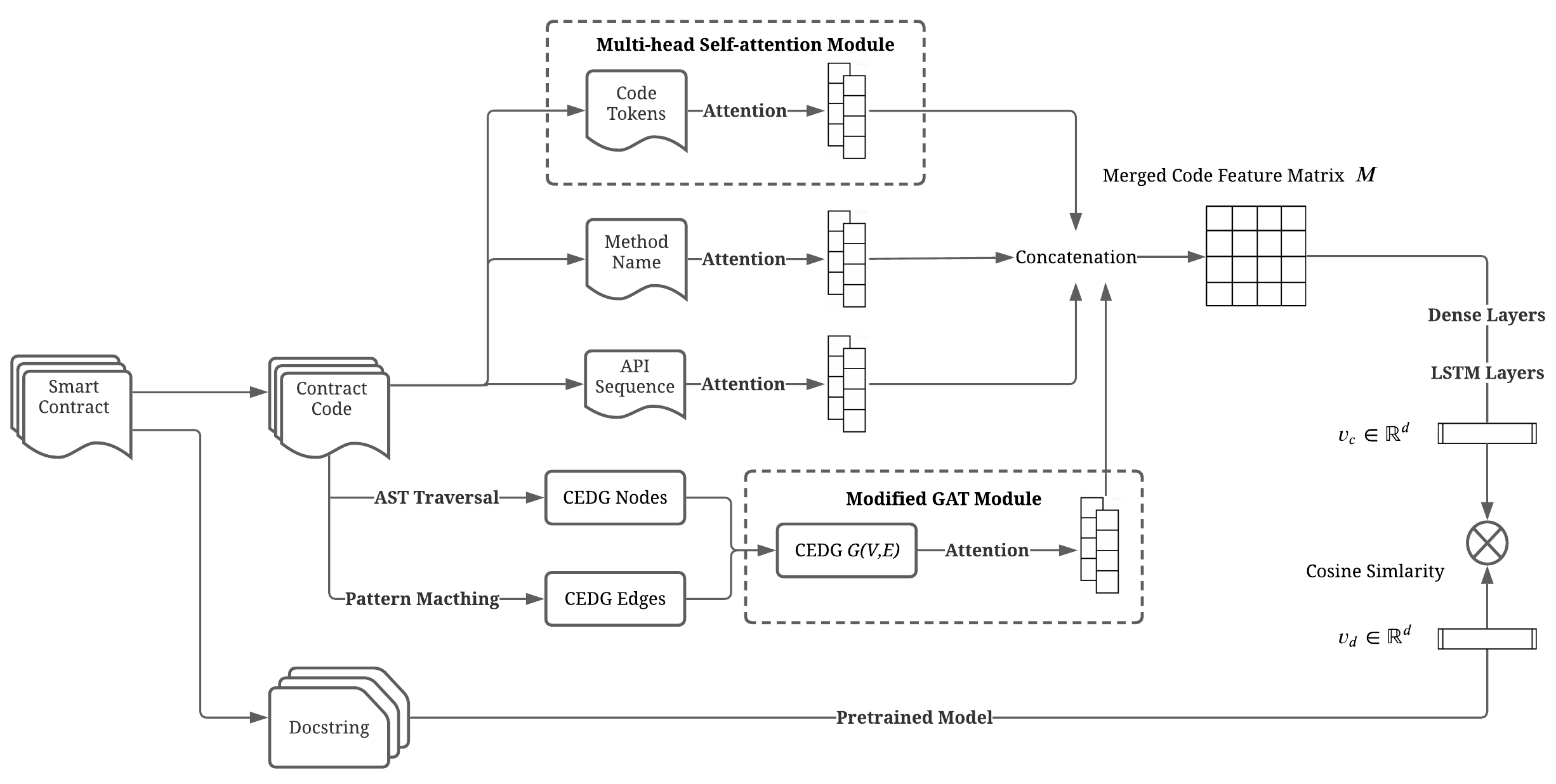}}
\caption{Overall Framework of MM-SCS.}
\label{fig3}
\end{figure*}

With the general architecture of neural code search in Fig.~\ref{fig1} as the backbone of MM-SCS, the overall architecture is shown in Fig.~\ref{fig3}. The embedding $v_c$ of the code snippet (at function level) is the concatenation of the embeddings of code tokens, function names, API sequence, and CEDG. Query embedding $v_q$ is substituted by the embedding $v_d$ of docstring (i.e. code comment) at the training stage. The main modules in MM-SCS are outlined as below:
\begin{itemize}
\item Three multi-head self-attention modules used to embed the textual information of code, including code tokens, function name, and API sequence.
\item One modified GAT module used to embed the CEDG generated from the code.
\item One pretrained module used to embed query information.
\item One neural network including LSTM and Dense layers used to output $v_c$ to the vector space of the same dimension as $v_d$.
\end{itemize}
These modules are applied respectively in the code embedding and query embedding processes. The subsections of this section will provide a detailed description of these modules as well as the training process.
\subsection{Code Embedding}
The input of code embedding falls into textual information and graph information. Textual information includes code tokens, function name, and API sequence; graph information stems from CEDG. The objective of code embedding is to map these inputs into a vector $v_c$ that can represent the entire code snippet.
\subsubsection{Embedding for Code Textual Information}
Emulating DeepCS, we extracted the textual information of code as $C_{text}=[T,F,A]$ via regular expressions. Where, $T=t_1,…,t_{l_T}$ of length $l_T$ is the code token sequence without special symbols. Each token written in camel-case or snake-case shall be segmented into the original word. For example, the token \textsf{itemCount} or \textsf{item\_count} shall be segmented into two separate words: item and count. $F=n_1,…,n_{l_F}$ of length $l_F$ and $A=a_1,…,a_{l_A}$ of length $l_A$ are the function name sequence and the API invocation sequence, respectively, which both are extracted from $T$.

\begin{figure}[htbp]
\centerline{\includegraphics[width=0.5\textwidth]{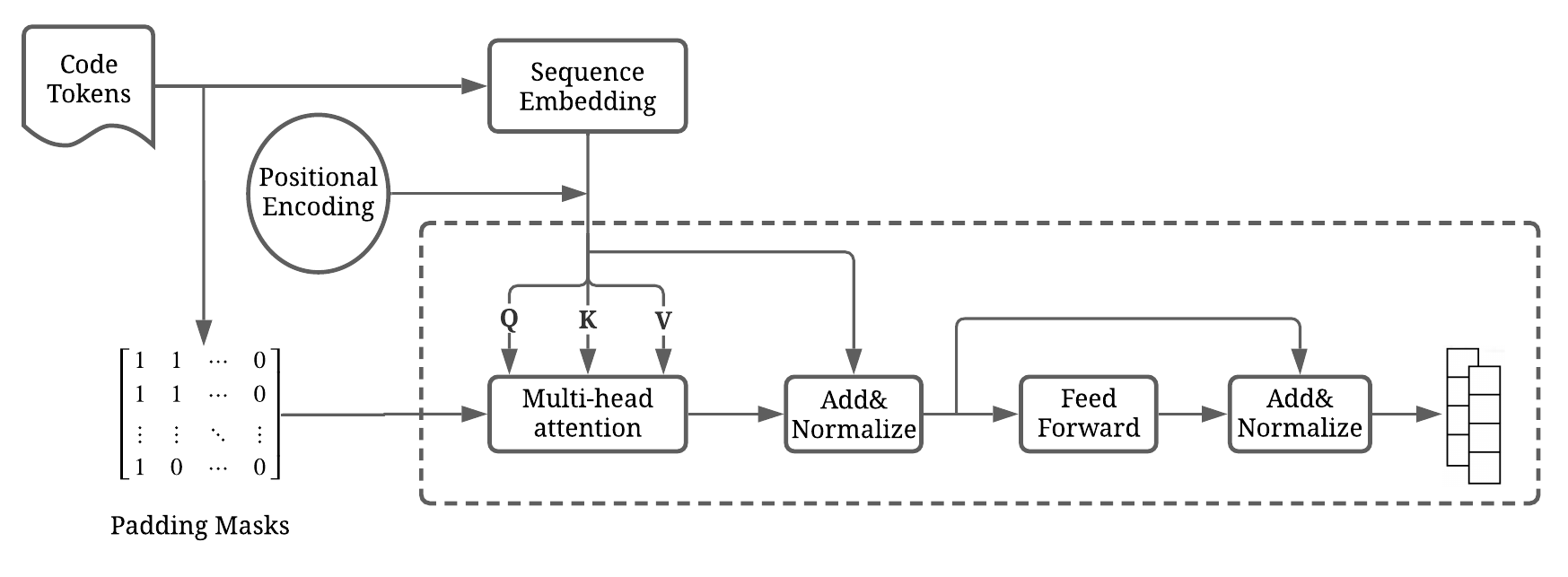}}
\caption{Multi-head Self-attention module.}
\label{fig4}
\end{figure}

To make MM-SCS better capture key contextual information, we use three multi-head self-attention modules to embed $T$, $F$ and $A$, respectively. The structure of multi-head self-attention module is shown in Fig~\ref{fig4}. It consists of a two-layer network made up of a multi-head self-attention and a position-wise feed-forward network. Each layer has an Add \& Norm sublayer, while Add denotes Residual Connection~\cite{DBLP:conf/cvpr/HeZRS16} which is used to prevent network degeneration, Norm denotes Layer Normalization~\cite{DBLP:journals/corr/BaKH16} which is used to normalize the activation value of each layer. Given that the embedding modules for $T$, $F$ and $A$ differ only in parameters, we here take the embedding process for $T$ as an example.

Similar to Transformer~\cite{DBLP:conf/nips/VaswaniSPUJGKP17}, we use positional encoding $PE$ to encode the relative positional information of words, as in Eq.~(\ref{eq6}). Where $pos$ denotes the position of a word in a sequence, $d$ denotes the dimension of $PE$ (the same as word embedding), and $i$ is the dimension; thus $2i$ denotes an even dimension and $2i+1$ denotes an odd dimension. In this way, the module can utilize the order information of sequences without recurrent or convolutional networks. By adding embedding and $PE$ of each word in $T$, we can get an input matrix $M_T\in R^{l_T\times d}$ for self-attention.
\begin{equation}
\begin{aligned}
\label{eq6}
PE_{(pos,2i)}=sin(pos/10000^{2i/d}),\\ PE_{(pos,2i+1)}=cos(pos/10000^{2i/d})
\end{aligned}
\end{equation}

Then, we derive the coefficients of attention $Q$, $K$ and $V$ by right-multiplying $M_T$ by three linear transformation matrices $W_Q\in R^{d\times d_k}$, $W_K\in R^{d\times d_k}$ and $W_V\in R^{d\times d_k}$, respectively, as in Eq.~(\ref{eq7}).

\begin{equation}
\begin{aligned}
\label{eq7}
Q=M_T  W_Q;~~K=M_T  W_K;~~V=M_T  W_V
\end{aligned}
\end{equation}

After that, we can calculate the output of self-attention as in Eq.~(\ref{eq8}), where $\sqrt{d_k}$ is the scaling factor to avoid gradient disappearing. Specifically, we multiply the input by the padding mask matrix prior to $softmax$ to shield the redundant information at the padding position. The multi-head self-attention network consists of multiple randomly initialized self-attention networks. Assume the number of heads is $J$, then the output $MulAtt$ of multi-head is as in Eq.~(\ref{eq9}).

\begin{equation}
\label{eq8}
Att(Q,K,V) = softmax(\frac{QK^\mathsf{T}}{\sqrt{d_k}})V
\end{equation}

\begin{equation}
\label{eq9}
MulAtt= Concat(Att_1,Att_2,\dots,Att_J)
\end{equation}

The value of $MulAtt$ of $T$ is input into the feed forward network after undergoing the Add \& Norm layer to get the final output $v_T$ as in Eq.~(\ref{eq10}), where \textbf{Relu} is the activation function, and $W_1$, $W_2$, $b_1$, $b_2$ are learnable parameters of the network. Likewise, we can derive the outputs, $v_F$ and $v_A$, of $F$ and $A$, respectively. 
\begin{equation}
\label{eq10}
v_T= \textbf{Relu}(MulAtt\cdot W_1+b_1)W_2+b_2
\end{equation}

\subsubsection{Embedding for CEDG}
\begin{figure}
\centerline{\includegraphics[width=0.5\textwidth]{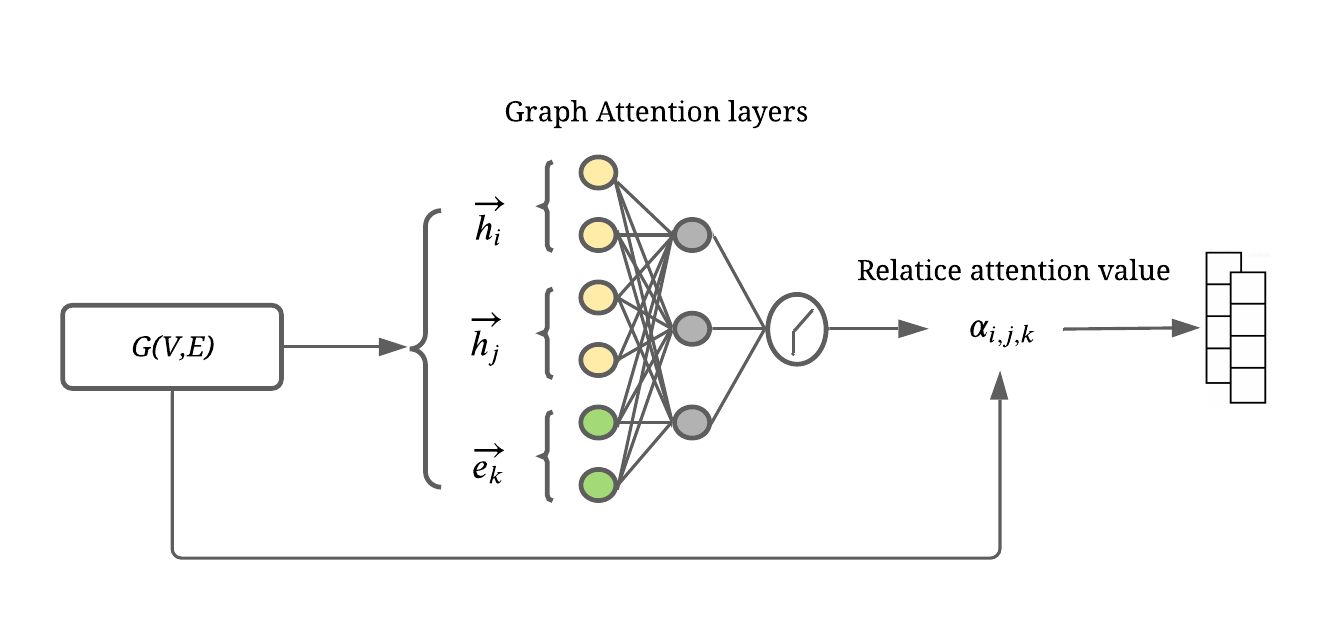}}
\caption{Modified GAT module.}
\label{fig5}
\end{figure}

The modified GAT module for CEDG embedding~\cite{DBLP:conf/acl/NathaniCSK19} is shown as Fig.~\ref{fig5}. $G(V,E)$ represents the CEDG constructed from code information. Split $G(V,E)$ into multiple triples $(\vec{h_i},\vec{h_j},\vec{e_k})$, where $\vec{h_i},\vec{h_j},\vec{e_k}$ are a vector representation for the head and tail nodes and for the relation (edge) connecting both, respectively. Multiply each triple by a linear transformation $W_1$ to learn the triple’s vector representation $\vec{c_{ijk}}$ as in Eq.~(\ref{eq11}).
\begin{equation}
\label{eq11}
\vec{c_{ijk}}= W_1[\vec{h_i}\|\vec{h_j}\|\vec{e_k}]
\end{equation}

Following this, we take $\vec{c_{ijk}}$ through a nonlinear layer $W_2$ activated by \textbf{LeakyRelu}, to get the parameter of attention weight, $b_{ijk}$, as in Eq.~(\ref{eq12}). Afterwards the attention weight $\alpha_{ijk}$ for each triple can be calculated as in Eq.~(\ref{eq13}).

\begin{equation}
\label{eq12}
b_{ijk}= \textbf{LeakyRelu}(W_2c_{ijk})
\end{equation}

\begin{equation}
\label{eq13}
\alpha_{ijk}= softmax_{jk}(b_{ijk})=\frac{exp(b_{ijk})}{\sum_{n\in N_i}\sum_{r\in R_{in}}exp(b_{ijk})}
\end{equation}

Where $N_i$ is the neighbor node to entity $i$, $R_{in}$ is the relation linking entity $i$ and $n$. The multi-head attention mechanism has also been included in this model to stabilize the learning process, so that more neighbor information can be learned. Finally, the new vector representation $\vec{h^{\prime}_i}$ for each node is calculated by Eq.~(\ref{eq14}).

\begin{equation}
\label{eq14}
\vec{h_i^\prime}= \frac{\sigma}{D}(\sum_{d=1}^D\sum_{j\in N_i}\sum_{k\in R_{ij}}\alpha_{ijk}^m\vec{c_{ijk}}^m)
\end{equation}
Where $D$ is the number of attention heads, and $\sigma$ is a nonlinearity. So far, the nodes of CEDG have learned the n-hop neighbor relationship between each other. By concatenating the vectors $\vec{h^{\prime}_i}$ for all the nodes of CEDG obtained after learning, we derive the vector representation $v_G$ for the whole CEDG.

\subsection{Embedding for Docstring(query)}
In this paper we use ALBERT~\cite{DBLP:conf/iclr/LanCGGSS20} model to embed docstring and query. ALBERT is a pretrained model on large-sized corpus in English, which is used to handle various downstream tasks of NLP. As the lite version of BERT, ALBERT possesses a less parameters and faster training speed at no sacrifice of effectiveness. Since docstring is also a natural language text in English, using the pretrained model can save training time and hardware resources and derive a more exact vector representation based on contexts with billions of sentences.

Although the considerably huge training data sets of ALBERT, its corpus stems mainly from books and English Wikipedia. The docstrings used to describe code snippets are slightly different from the sentences in these corpuses, as the former contains the terms that are related to blockchain and programming contexts. For instance, ``fallback'' refers specifically to a function type in docstring, carrying a completely different connotation from the ``fallback'' in general texts. To help ALBERT better understand blockchain contexts, we fine-tuned ALBERT using incremental training. Specifically, we performed n-gram masking for the docstrings in the training data sets and conducted training through sentence order prediction. ALBERT team provides customized scripts to help in this fine-tuning process~\cite{Albert}. Finally, we feed the docstrings into the fine-tuned ALBERT model to get a corresponding 768-dimensional vector representation $v_d$.
\subsection{Model Training}
As shown in Fig.~\ref{fig3}, we concatenate the code's multi-modal vector representations $v_T$,$v_F$,$v_A$,$v_G$ into a merged code feature matrix $M$ as in Eq.~(\ref{eq15}).
\begin{equation}
\label{eq15}
M=v_T \oplus v_F \oplus v_A \oplus v_G 
\end{equation}
The task of model optimization is to make code embedding $v_c$ as close as possible to the corresponding docstring embedding $v_d$ in a shared vector space. Therefore, we send $M$ to the LSTM layer, where it is further output and delivered to the dense layer, and finally the 768-dimensional code embedding $v_c$ is output. At the training stage, we provide a corresponding docstring embedding $v_d^+$ and a randomly chosen negative docstring embedding $v_d^-$ for each $v_c$. Our goal is to maximize the similarity between $v_c$ and $v_d^+$ and minimize the similarity between $v_c$ and $v_d^-$. Accordingly, the goal of training is to minimize the loss value $L(\theta)$ in Eq.~(\ref{eq2}). Where $\theta$ denotes the learnable parameters of the model, and $\beta$ is a constant margin, typically set as 0.05.
\begin{equation}
\label{eq2}
L(\theta) =max(0, \beta + cos(v_c,v_d^-) - cos(v_c,v_d^+) )
\end{equation} 

\subsection{Search Process}
After obtaining the well-trained MM-SCS model, we build a code repository for upcoming search tasks. We store the original code snippets and corresponding $v_c$ and set up indexes for them. When searching, MM-SCS embeds the query statement into vector $v_q$ using the fine-tuned ALBERT model. Since semantic similarity was measured in terms of the cosine distance between vectors as Eq.~(\ref{eq1}), MM-SCS searchs the most neighboring $v_c$ to $v_q$ in the vector space using Nmslib~\cite{nmslib} tool and returns the corresponding top-k rank. Finally, we can extract the detailed information about the corresponding code snippet as the search result via the index value of returned $v_c$.

\section{Evaluation}
All the experiments in this paper are implemented with Python 3.7, and run on Google Colab with a NVIDIA Tesla P100 GPU. The number of attention heads of both multi-head attention module and modified GAT module is set to 8, which is the widely used default setting in many tasks. Experiments investigate the following research questions (RQs):

\textbf{RQ1:} How is the effectiveness of MM-SCS compared with the state-of-the-art models?

\textbf{RQ2:} How are the training and operating efficiency of MM-SCS?

\textbf{RQ3:} How is the advantage of using CEDG as an extra modality?

\textbf{RQ4:} How much does the structural design affect the effectiveness of MM-SCS?
\subsection{Datasets and Baselines}
To build training and testing datasets, we collected 400k code snippets from Github and Etherscan.io using web clawers. After removing duplicate entries, there are 48,578 snippets remained in the dataset, including normal functions, modifiers and fallback functions. We segmented the code snippets into (code, docstring) pairs using regular expressions and conducted experiments using 10-fold cross validation method. Emulating baseline models, we used docstrings to simulate real queries, since docstrings are easily accessible and quite similar to queries in expression. 

We take four state-of-the-art models mentioned in Section 2.1: UNIF, DeepCS, CARLCS-CNN, and TAB-CS, as baselines for a comparison with the MM-SCS proposed herein. Besides the structural difference, the input modalities of models are also different, as shown in Table 3.
\begin{table}[]
\centering
\label{tab4}
\caption{Input Modalities of Baselines. $T$ is Code Token Sequence, $F$ is Function Name, $A$ is API Sequence}
\begin{tabular}{lccccc}
\hline
Model & $T$ & $F$ & $A$ & AST & CEDG \\ \hline
UNIF       & \checkmark           &             &            &    &     \\
DeepCS     & \checkmark             & \checkmark               & \checkmark              &    &     \\
CARLCS-CNN & \checkmark             & \checkmark               & \checkmark              &    &     \\
TAB-CS     & \checkmark            & \checkmark               & \checkmark              & \checkmark     &     \\
MM-SCS     & \checkmark             & \checkmark               & \checkmark              &    & \checkmark      \\ \hline
\end{tabular}\\
\end{table}

\subsection{Evaluation Metrics }
We used two metrics widely applied in code search tasks, SuccessRate@k~\cite{DBLP:conf/sigsoft/LiWWYXM16} (also referred to as Recall@k) and MRR~\cite{DBLP:conf/kbse/LvZLWZZ15, DBLP:conf/sigsoft/YeBL14}, to evaluate the effectiveness of MM-SCS.

\textbf{SuccessRate@k (SR@k)}: The percentage of queries for which can hit the standard answer among the top-k results of a single search. It is obtained by calculation from Eq.~(\ref{eq16}).
\begin{equation}
\label{eq16}
SR@k = \frac{1}{\left|Q\right|}\sum_{i=1}^{\left|Q\right|}\delta(Q_i<=k)
\end{equation}

Where $Q$ is a set of queries; $\delta()$ is an indicator function which returns 1 if the $i$--th query $Q_i$ has been hit successfully among the top-k results, else 0. The higher the SR@k value of the model, the better the effectiveness. Following the general settings in the experiment of baselines, we calculate SR values at $k$=1, 5, and 10.

\textbf{MRR}: The mean of reciprocal ranks of all queries, obtained by calculating from Eq.~(\ref{eq17}).

\begin{equation}
\label{eq17}
MRR = \frac{1}{\left|Q\right|}\sum_{i=1}^{\left|Q\right|}\frac{1}{Rank_{Q_i}}
\end{equation}

Where $Rank_{Q_i}$ refers to the rank of the first hit result of $Q_i$ among all search results. If there is no code snippet corresponding to $Q_i$ among the top-k results, then let $\frac{1}{Rank_{Q_i}}=0$. The higher the value of MRR, the more the result of the standard answer returned by the search engine comes near to the top. Likewise, following the general settings in baselines, we calculate MRR values at $k$=10.

\subsection{Experimental Results}
This section renders the experimental result for answering the RQs.
\subsubsection{RQ1: How is the effectiveness of MM-SCS compared with the state-of-the-art models?}
Table 4 compares the effectivenesses of MM-SCS and baselines on our dataset. MM-SCS achieves an MRR of 0.572, and SR@1/5/10 with 0.568/0.746/0.798. As compared to UNIF, DeepCS, CARLCS-CNN, and TAB-CS, MM-SCS has improved by 34.2\%, 59.3\%, 36.8\%, and 14.1\% in terms of MRR; by 36.2\%/45.1\%/40.7\%, 63.6\%/61.4\%/59.2\%, 30.8\%/36.8\%39.2\%, 16.1\%/19.7\\\%/20.1\% in terms of SR@1\%/5\%/10\%, respectively. Moreover, we have applied Wilcoxon signed-rank test~\cite{wilcoxon1992individual} at a 5\% significance level on all metrics between models, to get the p-values\textless 0.05. That means the experimental data corresponding to each model comes from entities with different distributions, proving the effectiveness advantage of MM-SCS versus baselines is statistically significant. 
\begin{table}[]
\centering
\label{tab5}
\caption{Effectiveness Comparison Between MM-SCS and Baselines in terms of SR@1/5/10 and MRR}
\begin{tabular}{lcccc}
\hline
Model      & SR@1 & SR@5 & SR@10 & MRR \\ \hline
UNIF       & 0.417           & 0.514             & 0.567            & 0.426   \\
DeepCS     & 0.347           & 0.462             & 0.501            & 0.359   \\
CARLCS-CNN & 0.434           & 0.545             & 0.573            & 0.418   \\
TAB-CS     & 0.489           & 0.623             & 0.664            & 0.501   \\
MM-SCS     & \textbf{0.568}         & \textbf{0.746}             & \textbf{0.798}          &\textbf{0.572}    \\ \hline
\end{tabular}
\end{table}

\subsubsection{RQ2: How are the training and operating efficiency of MM-SCS?}
Table 5 compares the training and testing time cost of MM-SCS and baselines on our dataset.  The result indicates that UNIF has the highest training and query efficiency, mainly because it only receives unimodal code input, as shown in Table 3. The training of MM-SCS is most time-consuming (10.2 hours), much slower than all baselines. According to the statistics, MM-SCS has taken 9.8 hours on average to fine-tune ALBERT and 0.4 hours to train the code embedding modules. Although the fine-tune process is time-consuming, ALBERT performs fast in the test. the operating efficiency of MM-SCS is 0.34s/query, which is only second to UNIF (0.19s/query). One main reason is that MM-SCS uses the attention mechanism which allows parallel computation in GPUs, while the recurrent neural networks used in DeepCS and CARLCS-CNN can only perform serial computation. Another reason is that compared with AST used in TAB-CS, CEDG has a simpler structure, which makes graph embedding much easier. Besides, the pretrained model ALBERT used in MM-SCS is lite. The operating time in our experiments considers both embedding time and retrieving time. In real repositories with indexes, embeddings would be stored in advance, and the retrieving speed is usually less than a few milliseconds.

\begin{table}[]
\centering
\label{tab6}
\caption{Time Cost for Training and Testing of models}
\begin{tabular}{lcc}
\hline
Model      & Training & Testing \\ \hline
UNIF       & \textbf{0.3 hours}           & \textbf{0.19s/q}            \\
DeepCS     & 3.8 hours           & 1.01s/q             \\
CARLCS-CNN & 1.4 hours           & 0.56s/q            \\
TAB-CS     & 0.8 hours           & 0.45s/q             \\
MM-SCS     & 10.2 hours           & 0.34s/q             \\ \hline
\end{tabular}
\end{table}

\subsubsection{RQ3: How is the advantage of using CEDG as an extra modality?}
Table 6 compares the effect of different input modalities on MM-SCS performance. According to the experimental result, when the textual modalities of code are fixed as ($T$+$A$+$F$), MM-SCS can improve by 22.6\%, 28.8\%, 30.3\%, and 23.8\% in terms of SR@1, SR@5, SR@10 and MRR, respectively, with CEDG as the extra input modality. With AST as the extra input modality, MM-SCS can only improve by 12.7\%, 14.6\%, 17.1\%, and 14.7\% in terms of SR@1, SR@5, SR@10 and MRR, respectively. Besides, the promotion brought about with AST+CEDG as the extra input modality is roughly the same as when CEDG is used alone. One possible reason is that CEDG has included the useful features for semantic search in AST, and that targeted design has been made for smart contracts. Moreover, the operating efficiency of using ($T$+$A$+$F$+CEDG) is 0.34/query, which is 0.8s faster than using ($T$+$A$+$F$+AST) in testing. Therefore, we tend to use ($T$+$A$+$F$+CEDG) as the input modalities of code in the real application scenarios of MM-SCS.

\begin{table}[]
\centering
\label{tab7}
\caption{Performance of MM-SCS with Different Modalities}
\scalebox{0.95}{
\begin{tabular}{lccccc}
\hline
\multicolumn{1}{c}{Modalities} & SR@1 & \multicolumn{1}{l}{SR@5} & \multicolumn{1}{l}{SR@10} & \multicolumn{1}{l}{MRR} & \multicolumn{1}{l}{Testing}\\ \hline
T                              &0.449      & 0.548                         &0.595                           &0.453    &\textbf{0.24s/q}                      \\
T+A+F                          &0.463      & 0.579                         &0.612                           &0.462      &0.29s/q                    \\
T+A+F+AST                      &0.522      & 0.664                         &0.717                           &0.530      &0.42s/q                  \\
T+A+F+CEDG                     &0.568      &\textbf{0.746 }                         &\textbf{0.798 }                          &0.572         &0.34s/q                \\
T+A+F+AST+CEDG                 &\textbf{0.573}      &0.737                          &0.791                           &\textbf{0.581}       &0.56s/q                    \\ \hline
\end{tabular}}
\end{table}

\subsubsection{RQ4: How much does the structural design affect the effectiveness of MM-SCS?}
Table 7 integrates Tables 3, 4 and 6 to compare the performances of MM-SCS and baselines in SR@1, SR@5, SR@10 and MRR under the same input modalities. The target is to infer the effects of different models’ structure on effectiveness. The experimental result indicates MM-SCS has improved by 7.6\%, 6.6\%, 4.9\%, and 6.3\%, respectively, versus UNIF, when the input modality is $T$. When the input modalities are ($T$+$A$+$F$) for all, MM-SCS has improved by 33.4\%, 25.3\%, 22.1\%, 28.6\% and 6.6\%, 6.2\%, 6.8\%, 10.5\% versus DeepCS and CARLCS-CNN, respectively. When the input modalities are ($T$+$A$+$F$+AST) for all, MM-SCS has improved by 6.7\%, 6.5\%, 7.9\%, and 5.7\%, respectively, versus TAB-CST. Except for UNIF, MM-SCS also operates 0.72s/0.27s/0.03s faster than DeepCS, CARLCS-CNN, and TAB-CS per query under same input modalities, respectively. We can infer that given the same input modalities, the structure of MM-SCS outperforms the structure of baselines.

\begin{table}[]
\centering
\label{tab8}
\caption{Effectiveness Comparison between Baselines and MM-SCS Under the Same Input Modalities}

\begin{tabular}{lcccc}
\hline
\multicolumn{1}{c}{Model} & SR@1 & \multicolumn{1}{l}{SR@5} & \multicolumn{1}{l}{SR@10} & \multicolumn{1}{l}{MRR} \\ \hline
UNIF                      & 0.417    & 0.514                        & 0.567                         & 0.426                       \\
MM-SCS($T$)                 & \textbf{0.449}    & \textbf{0.548}                        & \textbf{0.595}                         & \textbf{0.453}                    \\ \hline
DeepCS                    & 0.347    & 0.462                        & 0.501                         & 0.359                       \\
CARLCS-CNN                & 0.434    & 0.545                        & 0.573                         & 0.418                        \\
MM-SCS($T$+$A$+$F$)             & \textbf{0.463}    & \textbf{0.579}                        & \textbf{0.612}                         & \textbf{0.462}                         \\ \hline
TAB-CS                    & 0.489    & 0.623                        & 0.664                         & 0.501                     \\
MM-SCS($T$+$A$+$F$+AST)         & \textbf{0.522}    & \textbf{0.664}                        & \textbf{0.717}                         & \textbf{0.530}                     \\ \hline
\end{tabular}
\end{table}

\subsection{Case Study}
Listing 1 shows the fist retrieved results of MM-SCS and baselines for query ``destroy tokens of a certain address''. MM-SCS returned the most relevant result even there are no common words between code and query, while baselines returned irrelevant results. Actually, the term ``destroy'' barely appears in our corpus, which means supervised learning-based query embedding modules can hardly learn how to embed ``destroy''. Thus baselines can only understand the remaining sequence ``tokens of a certain address'', which leads the irrelevant results. MM-SCS takes a fine-tuned pretrained model to embed queries. During the fine-tune process, MM-SCS understood the  the synonymous relationship between ``burn tokens'' and ``destroy tokens'' in the context of blockchain. It indicates MM-SCS has better understanding of blockchain terms and contexts than baselines.
\lstset{
    numbers=none,
}

\begin{lstlisting}[caption = First retrieved results of MM-SCS and baselines for query ``destroy tokens of a certain address''.][language=Solidity]
//Query: "destroy tokens of a certain address"

//MM-SCS's first retrieved result 
function burn(uint256 _value) onlyOwner public returns (bool success) {
        require(_balanceOf[_owner] >= _value);
        require(_totalSupply >= _value);
        _balanceOf[_owner] -= _value;
        _totalSupply -= _value;
        Burn(_owner, _value);
        return true;
    }

//UNIF's and DeepCS's first retrieved result
function sendTokens(address _to, uint _value) public onlyMinter validAddress(_to) {
    // logic of sending tokens to a certain address
    }
    
// CARLCS-CNN's and TAB-CS's first retrieved result
function createTokens (uint256 _value) {
    // logic of creating new tokens
    }  
\end{lstlisting}

\section{Conclusion and Future Work}
In this paper, we propose an MM-SCS model for semantic search of smart contract code. In contrast to existing approaches, we implement multi-head self-attention mechanism for code embedding, make MM-SCS pay more attention to relatively important semantics. We also put forward a novel graph representation of smart contract, CEDG, as an extra modality to explore the 
hidden semantics between core code elements. Moreover, we adopt the fine-tuned ALBERT model to generate embeddings for queries when training data is limited. By comparing MM-SCS and the other four state-of-the-art baselines on the dataset with 470K entries built by us, the experimental result indicates MM-SCS is more suitable than baselines for semantic code search tasks of smart contracts.

Although using docstrings to simulate queries is a widely used method, authentic query data is conducive to enhancing the generalization ability of the model. Thus, we will enrich the size of our data sets and add authentic (code, query) pairs as the ground truth in the future. Moreover, We will add more types of nodes and edges for CEDG to represent more informative dependency between elements, and allow for the variances between different versions of Solidity. 

\bibliographystyle{ACM-Reference-Format}
\bibliography{sample-base}


\begin{thebibliography}{32}


\ifx \showCODEN    \undefined \def \showCODEN     #1{\unskip}     \fi
\ifx \showDOI      \undefined \def \showDOI       #1{#1}\fi
\ifx \showISBNx    \undefined \def \showISBNx     #1{\unskip}     \fi
\ifx \showISBNxiii \undefined \def \showISBNxiii  #1{\unskip}     \fi
\ifx \showISSN     \undefined \def \showISSN      #1{\unskip}     \fi
\ifx \showLCCN     \undefined \def \showLCCN      #1{\unskip}     \fi
\ifx \shownote     \undefined \def \shownote      #1{#1}          \fi
\ifx \showarticletitle \undefined \def \showarticletitle #1{#1}   \fi
\ifx \showURL      \undefined \def \showURL       {\relax}        \fi
\providecommand\bibfield[2]{#2}
\providecommand\bibinfo[2]{#2}
\providecommand\natexlab[1]{#1}
\providecommand\showeprint[2][]{arXiv:#2}

\bibitem[\protect\citeauthoryear{??}{Alb}{[n.\,d.]}]%
        {Albert}
 \bibinfo{year}{[n.\,d.]}\natexlab{}.
\newblock \bibinfo{title}{ALBERT}.
\newblock
\newblock
\urldef\tempurl%
\url{https://github.com/google-research/albert}
\showURL{%
\tempurl}
\newblock
\shownote{(accessed 10 May 2021)}.


\bibitem[\protect\citeauthoryear{??}{nms}{[n.\,d.]}]%
        {nmslib}
 \bibinfo{year}{[n.\,d.]}\natexlab{}.
\newblock \bibinfo{title}{Nmslib}.
\newblock
\newblock
\urldef\tempurl%
\url{https://github.com/nmslib/nmslib}
\showURL{%
\tempurl}
\newblock
\shownote{(accessed 10 May 2021)}.


\bibitem[\protect\citeauthoryear{??}{Sol}{[n.\,d.]}]%
        {Sol}
 \bibinfo{year}{[n.\,d.]}\natexlab{}.
\newblock \bibinfo{title}{Solidity 0.8.4 Documentation}.
\newblock
\newblock
\urldef\tempurl%
\url{https://docs.soliditylang.org/en/v0.8.4/}
\showURL{%
\tempurl}
\newblock
\shownote{(accessed 10 May 2021)}.


\bibitem[\protect\citeauthoryear{Ba, Kiros, and Hinton}{Ba
  et~al\mbox{.}}{2016}]%
        {DBLP:journals/corr/BaKH16}
\bibfield{author}{\bibinfo{person}{Lei~Jimmy Ba}, \bibinfo{person}{Jamie~Ryan
  Kiros}, {and} \bibinfo{person}{Geoffrey~E. Hinton}.}
  \bibinfo{year}{2016}\natexlab{}.
\newblock \showarticletitle{Layer Normalization}.
\newblock \bibinfo{journal}{\emph{CoRR}}  \bibinfo{volume}{abs/1607.06450}
  (\bibinfo{year}{2016}).
\newblock


\bibitem[\protect\citeauthoryear{Ben{-}Nun, Jakobovits, and Hoefler}{Ben{-}Nun
  et~al\mbox{.}}{2018}]%
        {DBLP:conf/nips/Ben-NunJH18}
\bibfield{author}{\bibinfo{person}{Tal Ben{-}Nun},
  \bibinfo{person}{Alice~Shoshana Jakobovits}, {and} \bibinfo{person}{Torsten
  Hoefler}.} \bibinfo{year}{2018}\natexlab{}.
\newblock \showarticletitle{Neural Code Comprehension: {A} Learnable
  Representation of Code Semantics}. In \bibinfo{booktitle}{\emph{NeurIPS}}.
  \bibinfo{pages}{3589--3601}.
\newblock


\bibitem[\protect\citeauthoryear{Bojanowski, Grave, Joulin, and
  Mikolov}{Bojanowski et~al\mbox{.}}{2017}]%
        {DBLP:journals/tacl/BojanowskiGJM17}
\bibfield{author}{\bibinfo{person}{Piotr Bojanowski}, \bibinfo{person}{Edouard
  Grave}, \bibinfo{person}{Armand Joulin}, {and} \bibinfo{person}{Tom{\'{a}}s
  Mikolov}.} \bibinfo{year}{2017}\natexlab{}.
\newblock \showarticletitle{Enriching Word Vectors with Subword Information}.
\newblock \bibinfo{journal}{\emph{Trans. Assoc. Comput. Linguistics}}
  \bibinfo{volume}{5} (\bibinfo{year}{2017}), \bibinfo{pages}{135--146}.
\newblock


\bibitem[\protect\citeauthoryear{Cambronero, Li, Kim, Sen, and
  Chandra}{Cambronero et~al\mbox{.}}{2019}]%
        {DBLP:conf/sigsoft/CambroneroLKS019}
\bibfield{author}{\bibinfo{person}{Jos{\'{e}} Cambronero},
  \bibinfo{person}{Hongyu Li}, \bibinfo{person}{Seohyun Kim},
  \bibinfo{person}{Koushik Sen}, {and} \bibinfo{person}{Satish Chandra}.}
  \bibinfo{year}{2019}\natexlab{}.
\newblock \showarticletitle{When deep learning met code search}. In
  \bibinfo{booktitle}{\emph{{ESEC/SIGSOFT} {FSE}}}. \bibinfo{publisher}{{ACM}},
  \bibinfo{pages}{964--974}.
\newblock


\bibitem[\protect\citeauthoryear{Can}{Can}{1993}]%
        {DBLP:journals/sigir/Can93}
\bibfield{author}{\bibinfo{person}{Fazli Can}.}
  \bibinfo{year}{1993}\natexlab{}.
\newblock \showarticletitle{Information Retrieval Data Structures {\&}
  Algorithms, by William B. Frakes and Ricardo Baeza-Yates (Book Review)}.
\newblock \bibinfo{journal}{\emph{{SIGIR} Forum}} \bibinfo{volume}{27},
  \bibinfo{number}{3} (\bibinfo{year}{1993}), \bibinfo{pages}{24--25}.
\newblock


\bibitem[\protect\citeauthoryear{Grover and Leskovec}{Grover and
  Leskovec}{2016}]%
        {DBLP:conf/kdd/GroverL16}
\bibfield{author}{\bibinfo{person}{Aditya Grover} {and} \bibinfo{person}{Jure
  Leskovec}.} \bibinfo{year}{2016}\natexlab{}.
\newblock \showarticletitle{node2vec: Scalable Feature Learning for Networks}.
  In \bibinfo{booktitle}{\emph{{KDD}}}. \bibinfo{publisher}{{ACM}},
  \bibinfo{pages}{855--864}.
\newblock


\bibitem[\protect\citeauthoryear{Gu, Zhang, and Kim}{Gu et~al\mbox{.}}{2018}]%
        {DBLP:conf/icse/GuZ018}
\bibfield{author}{\bibinfo{person}{Xiaodong Gu}, \bibinfo{person}{Hongyu
  Zhang}, {and} \bibinfo{person}{Sunghun Kim}.}
  \bibinfo{year}{2018}\natexlab{}.
\newblock \showarticletitle{Deep code search}. In
  \bibinfo{booktitle}{\emph{{ICSE}}}. \bibinfo{publisher}{{ACM}},
  \bibinfo{pages}{933--944}.
\newblock


\bibitem[\protect\citeauthoryear{He, Zhang, Ren, and Sun}{He
  et~al\mbox{.}}{2016}]%
        {DBLP:conf/cvpr/HeZRS16}
\bibfield{author}{\bibinfo{person}{Kaiming He}, \bibinfo{person}{Xiangyu
  Zhang}, \bibinfo{person}{Shaoqing Ren}, {and} \bibinfo{person}{Jian Sun}.}
  \bibinfo{year}{2016}\natexlab{}.
\newblock \showarticletitle{Deep Residual Learning for Image Recognition}. In
  \bibinfo{booktitle}{\emph{{CVPR}}}. \bibinfo{publisher}{{IEEE} Computer
  Society}, \bibinfo{pages}{770--778}.
\newblock


\bibitem[\protect\citeauthoryear{Hochreiter and Schmidhuber}{Hochreiter and
  Schmidhuber}{1997}]%
        {hochreiter1997long}
\bibfield{author}{\bibinfo{person}{Sepp Hochreiter} {and}
  \bibinfo{person}{J{\"u}rgen Schmidhuber}.} \bibinfo{year}{1997}\natexlab{}.
\newblock \showarticletitle{Long short-term memory}.
\newblock \bibinfo{journal}{\emph{Neural computation}} \bibinfo{volume}{9},
  \bibinfo{number}{8} (\bibinfo{year}{1997}), \bibinfo{pages}{1735--1780}.
\newblock


\bibitem[\protect\citeauthoryear{Husain, Wu, Gazit, Allamanis, and
  Brockschmidt}{Husain et~al\mbox{.}}{2019a}]%
        {DBLP:journals/corr/abs-1909-09436}
\bibfield{author}{\bibinfo{person}{Hamel Husain}, \bibinfo{person}{Ho{-}Hsiang
  Wu}, \bibinfo{person}{Tiferet Gazit}, \bibinfo{person}{Miltiadis Allamanis},
  {and} \bibinfo{person}{Marc Brockschmidt}.} \bibinfo{year}{2019}\natexlab{a}.
\newblock \showarticletitle{CodeSearchNet Challenge: Evaluating the State of
  Semantic Code Search}.
\newblock \bibinfo{journal}{\emph{CoRR}}  \bibinfo{volume}{abs/1909.09436}
  (\bibinfo{year}{2019}).
\newblock


\bibitem[\protect\citeauthoryear{Husain, Wu, Gazit, Allamanis, and
  Brockschmidt}{Husain et~al\mbox{.}}{2019b}]%
        {husain2019codesearchnet}
\bibfield{author}{\bibinfo{person}{Hamel Husain}, \bibinfo{person}{Ho-Hsiang
  Wu}, \bibinfo{person}{Tiferet Gazit}, \bibinfo{person}{Miltiadis Allamanis},
  {and} \bibinfo{person}{Marc Brockschmidt}.} \bibinfo{year}{2019}\natexlab{b}.
\newblock \showarticletitle{Codesearchnet challenge: Evaluating the state of
  semantic code search}.
\newblock \bibinfo{journal}{\emph{arXiv preprint arXiv:1909.09436}}
  (\bibinfo{year}{2019}).
\newblock


\bibitem[\protect\citeauthoryear{Kipf and Welling}{Kipf and Welling}{2017}]%
        {DBLP:conf/iclr/KipfW17}
\bibfield{author}{\bibinfo{person}{Thomas~N. Kipf} {and} \bibinfo{person}{Max
  Welling}.} \bibinfo{year}{2017}\natexlab{}.
\newblock \showarticletitle{Semi-Supervised Classification with Graph
  Convolutional Networks}. In \bibinfo{booktitle}{\emph{{ICLR} (Poster)}}.
  \bibinfo{publisher}{OpenReview.net}.
\newblock


\bibitem[\protect\citeauthoryear{Lan, Chen, Goodman, Gimpel, Sharma, and
  Soricut}{Lan et~al\mbox{.}}{2020}]%
        {DBLP:conf/iclr/LanCGGSS20}
\bibfield{author}{\bibinfo{person}{Zhenzhong Lan}, \bibinfo{person}{Mingda
  Chen}, \bibinfo{person}{Sebastian Goodman}, \bibinfo{person}{Kevin Gimpel},
  \bibinfo{person}{Piyush Sharma}, {and} \bibinfo{person}{Radu Soricut}.}
  \bibinfo{year}{2020}\natexlab{}.
\newblock \showarticletitle{{ALBERT:} {A} Lite {BERT} for Self-supervised
  Learning of Language Representations}. In \bibinfo{booktitle}{\emph{{ICLR}}}.
  \bibinfo{publisher}{OpenReview.net}.
\newblock


\bibitem[\protect\citeauthoryear{Li, Wang, Wang, Yan, Xie, and Mei}{Li
  et~al\mbox{.}}{2016}]%
        {DBLP:conf/sigsoft/LiWWYXM16}
\bibfield{author}{\bibinfo{person}{Xuan Li}, \bibinfo{person}{Zerui Wang},
  \bibinfo{person}{Qianxiang Wang}, \bibinfo{person}{Shoumeng Yan},
  \bibinfo{person}{Tao Xie}, {and} \bibinfo{person}{Hong Mei}.}
  \bibinfo{year}{2016}\natexlab{}.
\newblock \showarticletitle{Relationship-aware code search for JavaScript
  frameworks}. In \bibinfo{booktitle}{\emph{{SIGSOFT} {FSE}}}.
  \bibinfo{publisher}{{ACM}}, \bibinfo{pages}{690--701}.
\newblock


\bibitem[\protect\citeauthoryear{Lv, Zhang, Lou, Wang, Zhang, and Zhao}{Lv
  et~al\mbox{.}}{2015}]%
        {DBLP:conf/kbse/LvZLWZZ15}
\bibfield{author}{\bibinfo{person}{Fei Lv}, \bibinfo{person}{Hongyu Zhang},
  \bibinfo{person}{Jian{-}Guang Lou}, \bibinfo{person}{Shaowei Wang},
  \bibinfo{person}{Dongmei Zhang}, {and} \bibinfo{person}{Jianjun Zhao}.}
  \bibinfo{year}{2015}\natexlab{}.
\newblock \showarticletitle{CodeHow: Effective Code Search Based on {API}
  Understanding and Extended Boolean Model {(E)}}. In
  \bibinfo{booktitle}{\emph{{ASE}}}. \bibinfo{publisher}{{IEEE} Computer
  Society}, \bibinfo{pages}{260--270}.
\newblock


\bibitem[\protect\citeauthoryear{Mikolov, Chen, Corrado, and Dean}{Mikolov
  et~al\mbox{.}}{2013}]%
        {DBLP:journals/corr/abs-1301-3781}
\bibfield{author}{\bibinfo{person}{Tom{\'{a}}s Mikolov}, \bibinfo{person}{Kai
  Chen}, \bibinfo{person}{Greg Corrado}, {and} \bibinfo{person}{Jeffrey Dean}.}
  \bibinfo{year}{2013}\natexlab{}.
\newblock \showarticletitle{Efficient Estimation of Word Representations in
  Vector Space}. In \bibinfo{booktitle}{\emph{{ICLR} (Workshop Poster)}}.
\newblock


\bibitem[\protect\citeauthoryear{Nathani, Chauhan, Sharma, and Kaul}{Nathani
  et~al\mbox{.}}{2019}]%
        {DBLP:conf/acl/NathaniCSK19}
\bibfield{author}{\bibinfo{person}{Deepak Nathani}, \bibinfo{person}{Jatin
  Chauhan}, \bibinfo{person}{Charu Sharma}, {and} \bibinfo{person}{Manohar
  Kaul}.} \bibinfo{year}{2019}\natexlab{}.
\newblock \showarticletitle{Learning Attention-based Embeddings for Relation
  Prediction in Knowledge Graphs}. In \bibinfo{booktitle}{\emph{{ACL} {(1)}}}.
  \bibinfo{publisher}{Association for Computational Linguistics},
  \bibinfo{pages}{4710--4723}.
\newblock


\bibitem[\protect\citeauthoryear{Palangi, Deng, Shen, Gao, He, Chen, Song, and
  Ward}{Palangi et~al\mbox{.}}{2015}]%
        {DBLP:journals/corr/PalangiDSGHCSW15}
\bibfield{author}{\bibinfo{person}{Hamid Palangi}, \bibinfo{person}{Li Deng},
  \bibinfo{person}{Yelong Shen}, \bibinfo{person}{Jianfeng Gao},
  \bibinfo{person}{Xiaodong He}, \bibinfo{person}{Jianshu Chen},
  \bibinfo{person}{Xinying Song}, {and} \bibinfo{person}{Rabab~K. Ward}.}
  \bibinfo{year}{2015}\natexlab{}.
\newblock \showarticletitle{Deep Sentence Embedding Using the Long Short Term
  Memory Network: Analysis and Application to Information Retrieval}.
\newblock \bibinfo{journal}{\emph{CoRR}}  \bibinfo{volume}{abs/1502.06922}
  (\bibinfo{year}{2015}).
\newblock


\bibitem[\protect\citeauthoryear{Perozzi, Al{-}Rfou, and Skiena}{Perozzi
  et~al\mbox{.}}{2014}]%
        {DBLP:conf/kdd/PerozziAS14}
\bibfield{author}{\bibinfo{person}{Bryan Perozzi}, \bibinfo{person}{Rami
  Al{-}Rfou}, {and} \bibinfo{person}{Steven Skiena}.}
  \bibinfo{year}{2014}\natexlab{}.
\newblock \showarticletitle{DeepWalk: online learning of social
  representations}. In \bibinfo{booktitle}{\emph{{KDD}}}.
  \bibinfo{publisher}{{ACM}}, \bibinfo{pages}{701--710}.
\newblock


\bibitem[\protect\citeauthoryear{Sachdev, Li, Luan, Kim, Sen, and
  Chandra}{Sachdev et~al\mbox{.}}{2018}]%
        {DBLP:conf/pldi/SachdevLLKS018}
\bibfield{author}{\bibinfo{person}{Saksham Sachdev}, \bibinfo{person}{Hongyu
  Li}, \bibinfo{person}{Sifei Luan}, \bibinfo{person}{Seohyun Kim},
  \bibinfo{person}{Koushik Sen}, {and} \bibinfo{person}{Satish Chandra}.}
  \bibinfo{year}{2018}\natexlab{}.
\newblock \showarticletitle{Retrieval on source code: a neural code search}. In
  \bibinfo{booktitle}{\emph{MAPL@PLDI}}. \bibinfo{publisher}{{ACM}},
  \bibinfo{pages}{31--41}.
\newblock


\bibitem[\protect\citeauthoryear{Shuai, Xu, Liu, Yan, Xia, and Lei}{Shuai
  et~al\mbox{.}}{2020}]%
        {DBLP:conf/iwpc/ShuaiX0Y0L20}
\bibfield{author}{\bibinfo{person}{Jianhang Shuai}, \bibinfo{person}{Ling Xu},
  \bibinfo{person}{Chao Liu}, \bibinfo{person}{Meng Yan}, \bibinfo{person}{Xin
  Xia}, {and} \bibinfo{person}{Yan Lei}.} \bibinfo{year}{2020}\natexlab{}.
\newblock \showarticletitle{Improving Code Search with Co-Attentive
  Representation Learning}. In \bibinfo{booktitle}{\emph{{ICPC}}}.
  \bibinfo{publisher}{{ACM}}, \bibinfo{pages}{196--207}.
\newblock


\bibitem[\protect\citeauthoryear{Vaswani, Shazeer, Parmar, Uszkoreit, Jones,
  Gomez, Kaiser, and Polosukhin}{Vaswani et~al\mbox{.}}{2017}]%
        {DBLP:conf/nips/VaswaniSPUJGKP17}
\bibfield{author}{\bibinfo{person}{Ashish Vaswani}, \bibinfo{person}{Noam
  Shazeer}, \bibinfo{person}{Niki Parmar}, \bibinfo{person}{Jakob Uszkoreit},
  \bibinfo{person}{Llion Jones}, \bibinfo{person}{Aidan~N. Gomez},
  \bibinfo{person}{Lukasz Kaiser}, {and} \bibinfo{person}{Illia Polosukhin}.}
  \bibinfo{year}{2017}\natexlab{}.
\newblock \showarticletitle{Attention is All you Need}. In
  \bibinfo{booktitle}{\emph{{NIPS}}}. \bibinfo{pages}{5998--6008}.
\newblock


\bibitem[\protect\citeauthoryear{Veli{\v{c}}kovi{\'c}, Cucurull, Casanova,
  Romero, Lio, and Bengio}{Veli{\v{c}}kovi{\'c} et~al\mbox{.}}{2017}]%
        {velivckovic2017graph}
\bibfield{author}{\bibinfo{person}{Petar Veli{\v{c}}kovi{\'c}},
  \bibinfo{person}{Guillem Cucurull}, \bibinfo{person}{Arantxa Casanova},
  \bibinfo{person}{Adriana Romero}, \bibinfo{person}{Pietro Lio}, {and}
  \bibinfo{person}{Yoshua Bengio}.} \bibinfo{year}{2017}\natexlab{}.
\newblock \showarticletitle{Graph attention networks}.
\newblock \bibinfo{journal}{\emph{arXiv preprint arXiv:1710.10903}}
  (\bibinfo{year}{2017}).
\newblock


\bibitem[\protect\citeauthoryear{Wang, Li, Ma, Xia, and Jin}{Wang
  et~al\mbox{.}}{2020}]%
        {wang2020detecting}
\bibfield{author}{\bibinfo{person}{Wenhan Wang}, \bibinfo{person}{Ge Li},
  \bibinfo{person}{Bo Ma}, \bibinfo{person}{Xin Xia}, {and}
  \bibinfo{person}{Zhi Jin}.} \bibinfo{year}{2020}\natexlab{}.
\newblock \showarticletitle{Detecting code clones with graph neural network and
  flow-augmented abstract syntax tree}. In \bibinfo{booktitle}{\emph{2020 IEEE
  27th International Conference on Software Analysis, Evolution and
  Reengineering (SANER)}}. IEEE, \bibinfo{pages}{261--271}.
\newblock


\bibitem[\protect\citeauthoryear{Wilcoxon}{Wilcoxon}{1992}]%
        {wilcoxon1992individual}
\bibfield{author}{\bibinfo{person}{Frank Wilcoxon}.}
  \bibinfo{year}{1992}\natexlab{}.
\newblock \showarticletitle{Individual comparisons by ranking methods}.
\newblock In \bibinfo{booktitle}{\emph{Breakthroughs in statistics}}.
  \bibinfo{publisher}{Springer}, \bibinfo{pages}{196--202}.
\newblock


\bibitem[\protect\citeauthoryear{Xu, Yang, Liu, Shuai, Yan, Lei, and Xu}{Xu
  et~al\mbox{.}}{2021}]%
        {DBLP:conf/wcre/XuYLSYL021}
\bibfield{author}{\bibinfo{person}{Ling Xu}, \bibinfo{person}{Huanhuan Yang},
  \bibinfo{person}{Chao Liu}, \bibinfo{person}{Jianhang Shuai},
  \bibinfo{person}{Meng Yan}, \bibinfo{person}{Yan Lei}, {and}
  \bibinfo{person}{Zhou Xu}.} \bibinfo{year}{2021}\natexlab{}.
\newblock \showarticletitle{Two-Stage Attention-Based Model for Code Search
  with Textual and Structural Features}. In
  \bibinfo{booktitle}{\emph{{SANER}}}. \bibinfo{publisher}{{IEEE}},
  \bibinfo{pages}{342--353}.
\newblock


\bibitem[\protect\citeauthoryear{Yamaguchi, Golde, Arp, and Rieck}{Yamaguchi
  et~al\mbox{.}}{2014}]%
        {DBLP:conf/sp/YamaguchiGAR14}
\bibfield{author}{\bibinfo{person}{Fabian Yamaguchi}, \bibinfo{person}{Nico
  Golde}, \bibinfo{person}{Daniel Arp}, {and} \bibinfo{person}{Konrad Rieck}.}
  \bibinfo{year}{2014}\natexlab{}.
\newblock \showarticletitle{Modeling and Discovering Vulnerabilities with Code
  Property Graphs}. In \bibinfo{booktitle}{\emph{{IEEE} Symposium on Security
  and Privacy}}. \bibinfo{publisher}{{IEEE} Computer Society},
  \bibinfo{pages}{590--604}.
\newblock


\bibitem[\protect\citeauthoryear{Ye, Bunescu, and Liu}{Ye
  et~al\mbox{.}}{2014}]%
        {DBLP:conf/sigsoft/YeBL14}
\bibfield{author}{\bibinfo{person}{Xin Ye}, \bibinfo{person}{Razvan~C.
  Bunescu}, {and} \bibinfo{person}{Chang Liu}.}
  \bibinfo{year}{2014}\natexlab{}.
\newblock \showarticletitle{Learning to rank relevant files for bug reports
  using domain knowledge}. In \bibinfo{booktitle}{\emph{{SIGSOFT} {FSE}}}.
  \bibinfo{publisher}{{ACM}}, \bibinfo{pages}{689--699}.
\newblock


\bibitem[\protect\citeauthoryear{Zhuang, Liu, Qian, Liu, Wang, and He}{Zhuang
  et~al\mbox{.}}{2020}]%
        {zhuang2020smart}
\bibfield{author}{\bibinfo{person}{Yuan Zhuang}, \bibinfo{person}{Zhenguang
  Liu}, \bibinfo{person}{Peng Qian}, \bibinfo{person}{Qi Liu},
  \bibinfo{person}{Xiang Wang}, {and} \bibinfo{person}{Qinming He}.}
  \bibinfo{year}{2020}\natexlab{}.
\newblock \showarticletitle{Smart Contract Vulnerability Detection using Graph
  Neural Network.}. In \bibinfo{booktitle}{\emph{IJCAI}}.
  \bibinfo{pages}{3283--3290}.
\newblock


\end{thebibliography}


\end{document}